\documentclass[12pt]{article}
\usepackage{epsfig}
\usepackage{subfigure}
\usepackage{amssymb,amsmath}
\usepackage{hyperref}
\usepackage{setspace}
\usepackage{url}

\def\tst{\tilde t}
\def\ttau{\tilde \tau}

\begin{document}


\vspace*{1.0cm}

\begin{center}
\baselineskip 20pt {\Large\bf Yukawa Unification and Neutralino Dark
Matter in $SU(4)_c \times SU(2)_L \times SU(2)_R$ }

\vspace{1cm}
{\large Ilia Gogoladze\footnote{ E-mail: ilia@physics.udel.edu\\
\hspace*{0.5cm} On  leave of absence from: Andronikashvili Institute
of Physics, GAS, 380077 Tbilisi, Georgia.} Rizwan Khalid\footnote{
E-mail: rizwan@udel.edu} and Qaisar Shafi  } \vspace{0.5cm}

{\baselineskip 20pt \it
$^a$Bartol Research Institute, Department of Physics and Astronomy, \\
 University of Delaware, Newark, DE 19716, USA \\
}
\vspace{.5cm}

 \vspace{1.5cm} {\bf Abstract}
\end{center}

We consider a left-right symmetric $SU(4)_c \times SU(2)_L \times
SU(2)_R$ (4-2-2) model with gravity mediated supersymmetry breaking.
We find that with 4-2-2 compatible non-universal gaugino masses,
$t-b-\tau$ Yukawa coupling unification is consistent
 with neutralino dark matter abundance and with constraints from
collider experiments (except $(g-2)_{\mu}$). The gluino mass lies
close to that of the lightest neutralino, so that  the gluino
co-annihilation channel plays an important role in determining the
neutralino relic abundance. By relaxing the Yukawa unification
constraint we find stau and stop masses as low as $200-220$ GeV. We
highlight some benchmark points for these cases with $40 \leq
\tan\beta \leq 58$.

\thispagestyle{empty}

\newpage

\addtocounter{page}{-1}

\baselineskip 18pt


\section{Introduction}
As a maximal subgroup of Spin(10) (commonly known as $SO(10)$), the
gauge symmetry $SU(4)_c \times SU(2)_L \times SU(2)_R$ (4-2-2)
\cite{pati} captures many salient features exhibited by its covering
group. Even as a stand alone symmetry group, 4-2-2 implements
electric charge quantization, albeit in units of $\pm e/6$, rather
than $\pm e/3$ \cite{Lazarides:1980tg,King:1997ia}. It explains the
standard model quantum numbers of the quark and lepton families by
assigning them in bi-fundamental representations and it also
predicts the existence of right handed neutrinos \cite{pati}.
However, there are some important differences between models based
on $SO(10)$ and 4-2-2 which, in principle, can be experimentally
tested. For instance, in 4-2-2 the lightest magnetic monopole
carries two quanta of Dirac magnetic charge \cite{Lazarides:1980cc}.
(In $SO(10)$ the lightest monopole carries one quantum of Dirac
magnetic charge, unless $SO(10)$ breaks via 4-2-2.) By the same
token, 4-2-2 predicts the existence of $SU(3)$ color singlet states
carrying electric charges $\pm e/2$
\cite{King:1997ia,Kephart:2006zd}. Finally, gauge boson mediated
proton decay is a characteristic feature of $SO(10)$ which is absent
in the 4-2-2 framework.

While these different experimental signatures can help distinguish
$SO(10)$ from 4-2-2, they mostly rely on physics operating at superheavy
scales. A major motivation for this paper is to highlight some important
differences in the low energy predictions of supersymmetric $SO(10)$ and
4-2-2 models, stemming from the Higgs and sparticle sectors of these
models. An exciting new feature is that these predictions can be
sufficiently different so that they can be compared at the LHC.

Supplementing 4-2-2 with a discrete left-right (LR) symmetry \cite{
pati,lr}(more precisely C-parity) \cite{c-parity} reduces from three
to two the number of independent gauge couplings in supersymmetric
4-2-2. In combination with Yukawa unification \cite{big-422}, this
has important implications for low energy Higgs and sparticle
spectroscopy which we will explore in this paper, and compare with
the corresponding predictions from an SO(10) model.

In 4-2-2 the matter fields are unified into three generations of
$\psi$ (4, 2, 1), and the antimatter
fields are in three generations of $\psi_c$ (4, 1, 2). If the MSSM
electroweak doublets come from the bi-doublet H(1, 2, 2), the third
family Yukawa coupling H $\psi_c \psi$ yields the following relation
valid at the GUT scale ($M_{\rm GUT}$), namely
\begin{equation}
Y_t = Y_b = Y_{\tau} = Y_{\rm Dirac}.
\end{equation}

We will assume that due to C-parity the soft mass$^2$ terms, induced
at $M_{GUT}$ through gravity mediated supersymmetry breaking
\cite{Chamseddine:1982jx}, are equal in magnitude for the scalar
squarks and leptons of the three families. The asymptotic MSSM
gaugino masses, on the other hand, can be non-universal from the
following consideration. From C-parity, we can expect that the
gaugino masses at $M_{GUT}$ associated with $SU(2)_L$ and $SU(2)_R$
are the same. However, the asymptotic $SU(4)_c$ and consequently
$SU(3)_c$ gaugino masses can be different. With the hypercharge
generator in 4-2-2 given by $Y=\sqrt{\frac{2}{5}}
(B-L)+\sqrt{\frac{3}{5}} I_{3R}$, where $B-L$ and $I_{3R}$ are the
diagonal generators of $SU(4)_c$ and $SU(2)_R$, we have the
following asymptotic relation between the three MSSM gaugino masses:
\begin{equation}
M_1=\frac{3}{5} M_2 + \frac{2}{5} M_3. \label{gaugino-condition}
\end{equation}
The supersymmetric 4-2-2 model with C-parity thus has two
independent parameters $(M_2,\, M_3)$ in the gaugino sector.

In this paper we wish to explore whether Yukawa coupling unification
in 4-2-2 is compatible with recent observations of the dark matter
relic abundance and other collider-based experimental constraints. A
similar analysis for $SO(10)$, which we closely follow, has been
carried out by Baer {\it et al.} \cite{Baer:2008jn}. Solutions
consistent with SO(10)Yukawa unification have been obtained in
\cite{Baer:2008jn} only for very special values of the fundamental
parameters. Furthermore, it turns out to be quite difficult in this
model to reconcile the lightest neutralino primordial abundance with
the observed dark matter densities.

 By introducing non-universality
in the gaugino sector, we can allow the neutralinos in 4-2-2 to be
closely degenerate in mass with the gluino, which is not possible in
$SO(10)$. This  opens up,  in particular, the bino-gluino
co-annihilation channel \cite{Profumo:2004wk}, which  turns out to
be an essential difference between the 4-2-2 and $SO(10)$ models. In
order to make Yukawa coupling unification compatible with radiative
electroweak symmetry breaking (REWSB), one needs to implement some
splitting in the Higgs sector, with ${m^2_{Hu}} < {m^2_{Hd}}$. Such
a splitting may be introduced via a $D$-term contribution to all
scalar masses \cite{dterms}, or it can be generated via GUT scale
threshold corrections related to a large Dirac neutrino Yukawa
coupling \cite{Blazek:2002ta}. It has been noted \cite{Auto:2003ys}
that a splitting just in the Higgs soft terms, as opposed to
splitting in all scalar masses, yields better Yukawa unification,
and so we focus on this approach. Since one of our goals is a
comparison of 4-2-2 and $SO(10)$ models, we follow the same notation
as in \cite{Baer:2008jn}. We parameterize the Higgs soft mass
splitting by $m^2_{H_{u,d}}=m_{10}^2\mp 2 M_D^2$, where $m^2_{10}$
is the MSSM
 universal Higgs soft mass$^2$ term. The supersymmetric 4-2-2 model
 we are discussing thus has the following fundamental parameters:
\begin{equation}
m_{16}, m_{10}, M_D, M_2, M_3, A_0, \tan\beta, {\rm sign}~{\mu}.
\end{equation}
Thus, compared to the $SO(10)$ model of \cite{Baer:2008jn}, we have
one additional parameter in 4-2-2 which plays a crucial role in
realizing  Yukawa unification consistent with the desired neutralino
relic density.

The outline for the rest of the paper is as follows. In Section
\ref{constraints_section} we summarize the scanning procedure and the
experimental constraints that we have employed. We present the results from our
scan in Section \ref{results}, where we compare the 4-2-2 and $SO(10)$ models
and then proceed to highlight some of the predictions of the 4-2-2 model.
Our conclusions are summarized in Section \ref{conclusions}.

\section{Phenomenological constraints and scanning procedure\label{constraints_section}} \label{ch:constraints}

We employ ISAJET~7.78 package~\cite{ISAJET} to perform random scans
over the parameter space. In this package, the weak scale values of
gauge and third generation Yukawa couplings are evolved to $M_{\rm
GUT}$ via the MSSM renormalization group equations (RGEs) in the
$\overline{DR}$ regularization scheme, where $M_{\rm GUT}$ is
defined to be the scale at which $g_1=g_2$. We do not enforce an
exact unification of the strong coupling $g_3=g_1=g_2$ at $M_{\rm
GUT}$, since a few percent deviation from unification can be
assigned to unknown GUT-scale threshold
corrections~\cite{Hisano:1992jj}. At $M_{\rm GUT}$, the boundary
conditions are imposed and all the SSB parameters, along with the
gauge and Yukawa couplings, are evolved back to the weak scale
$M_{\rm Z}$. The effect of the neutrino Dirac Yukawa coupling in the
running of the RGEs has been shown in \cite{Gomez:2009yc} to be
significant  for coupling values $\sim 2$. In the 4-2-2 model with
$t-b-\tau$ unification, the asymptotic neutrino Dirac Yukawa
coupling has the same value as $y_t {\rm (M_{GUT})}$ which is
relatively small ($\sim 0.5$). Thus, in the following discussion we
will ignore it.

In the evaluation of Yukawa couplings the SUSY threshold corrections~\cite{Pierce:1996zz}
are taken into account at the common scale $M_{\rm SUSY}= \sqrt{m_{\tst_L}m_{\tst_R}}$.
The entire parameter set is iteratively run between $M_{\rm Z}$ and $M_{\rm GUT}$
using the full 2-loop RGEs until a stable solution is obtained. To
better account for leading-log corrections, one-loop step-beta functions
are adopted for gauge and Yukawa couplings, and the SSB parameters $m_i$
are extracted from RGEs at multiple scales $m_i=m_i(m_i)$. The RGE-improved
1-loop effective potential is minimized at an optimized scale $M_{\rm SUSY}$,
which effectively accounts for the leading 2-loop corrections. Full 1-loop
radiative corrections are incorporated for all sparticle masses.

The requirement of radiative electroweak symmetry breaking (REWSB)~\cite{Ibanez:1982fr}
puts an important theoretical constraint on the parameter space. Another
important constraint comes from limits on the cosmological abundance of
stable charged particles~\cite{Yao:2006px}. This excludes regions in the
parameter space where charged SUSY particles, such as $\ttau_1$ or $\tst_1$,
become the lightest supersymmetric particle (LSP). We accept only those
solutions for which one of the neutralinos is the LSP.

We have performed random scans for the following parameter range:
\begin{eqnarray}
0\leq & m_{16} & \leq 20\, \rm{TeV}, \nonumber \\
0\leq & M_{2} & \leq 1\, \rm{TeV}, \nonumber  \\
0\leq & M_{3} & \leq 1\, \rm{TeV}, \nonumber  \\
-3 \leq & A_{0}/m_{16} & \leq 0, \nonumber \\
0 \leq & M_{D}/m_{16} & \leq 0.95 , \nonumber \\
0 \leq & m_{10}/m_{16} & \leq 1.5, \nonumber \\
40 \leq & \tan \beta & \leq 58,
\label{ppp1}
\end{eqnarray}
with $\mu >0$ , and $m_t = 172.6$~GeV \cite{Group:2008nq}.

We first collected 150,000 points for both the $SO(10)$ and 4-2-2 models.
All of these points satisfy the requirement of REWSB with the neutralino being the LSP in each case.
Furthermore, all of these points satisfy the constraint $\Omega_{\rm CDM}h^2 \le 10$.
This is done so as to collect more points with a WMAP compatible value of cold dark
matter relic abundance. Once we identify good regions in paramter space, we perform a random scan
focused around those regions for the 4-2-2 case. After collecting the data, we use the
IsaTools package~\cite{Baer:2002fv}
to implement the following phenomenological constraints:
\begin{eqnarray}
m_{\tilde{\chi}^{\pm}_{1}}~{\rm (chargino~mass)} \geq 103.5~{\rm GeV} \qquad  \cite{Yao:2006px},\nonumber \\
m_h~{\rm (lightest~Higgs~mass)} \geq 114.4~{\rm GeV} \qquad  \cite{Schael:2006cr},\nonumber \\
m_{\tilde \tau}~{\rm (stau~mass)} \geq 86~{\rm GeV} \qquad \cite{Yao:2006px},\nonumber \\
m_{\tilde g}~{\rm (gluino~mass)} \geq 220~{\rm GeV} \qquad \cite{Yao:2006px},\nonumber \\
BR(B_s \rightarrow \mu^+ \mu^-)< 5.8 \times 10^{-8} \qquad  \cite{:2007kv},\nonumber \\
2.85 \times 10^{-4} \leq BR(b \rightarrow s \gamma)\leq 4.24 \times 10^{-4} \; (2\sigma) \;  \cite{Barberio:2007cr},\nonumber \\
\Omega_{\rm CDM}h^2 = 0.111^{+0.028}_{-0.037} \;(5\sigma) \qquad \qquad  \cite{Komatsu:2008hk},\nonumber \\
3.4 \times 10^{-10}\leq \Delta a_{\mu} \leq 55.6 \times 10^{-10}~ \; (3\sigma) \qquad  \cite{Bennett:2006fi}.
\end{eqnarray}
We apply the experimental constraints successively on
the data that we acquire from ISAJET. As a first step we apply the constraints
from $BR(B_s\rightarrow \mu^+ \mu^-)$, $BR(b\rightarrow s \gamma)$, the WMAP
upper bound on the relic density of cold dark matter, and the (s)particle mass bounds.
We then apply the WMAP lower bound on the relic density of dark matter, followed
by the constraint on the muon anomalous
magnetic moment $a_{\mu}=(g-2)_{\mu}/2$ at the $3\sigma$ allowed region.
The data is then plotted showing the successive application of these constraints.

\section{Results \label{results}}

Following Baer {\it et al.} \cite{Baer:2008jn} we introduce a parameter $R$ to quantify Yukawa unification.
Namely, $R$ is the ratio,
\begin{eqnarray}
R=\frac{ {\rm max}(y_t,y_b,y_{\tau})}{ {\rm min}(y_t,y_b,y_{\tau})},
\end{eqnarray}
so that $R=1$ corresponds to perfect unification and a higher value of $R$ signifies
a larger deviation from unification.

\begin{figure}
\centering
\includegraphics{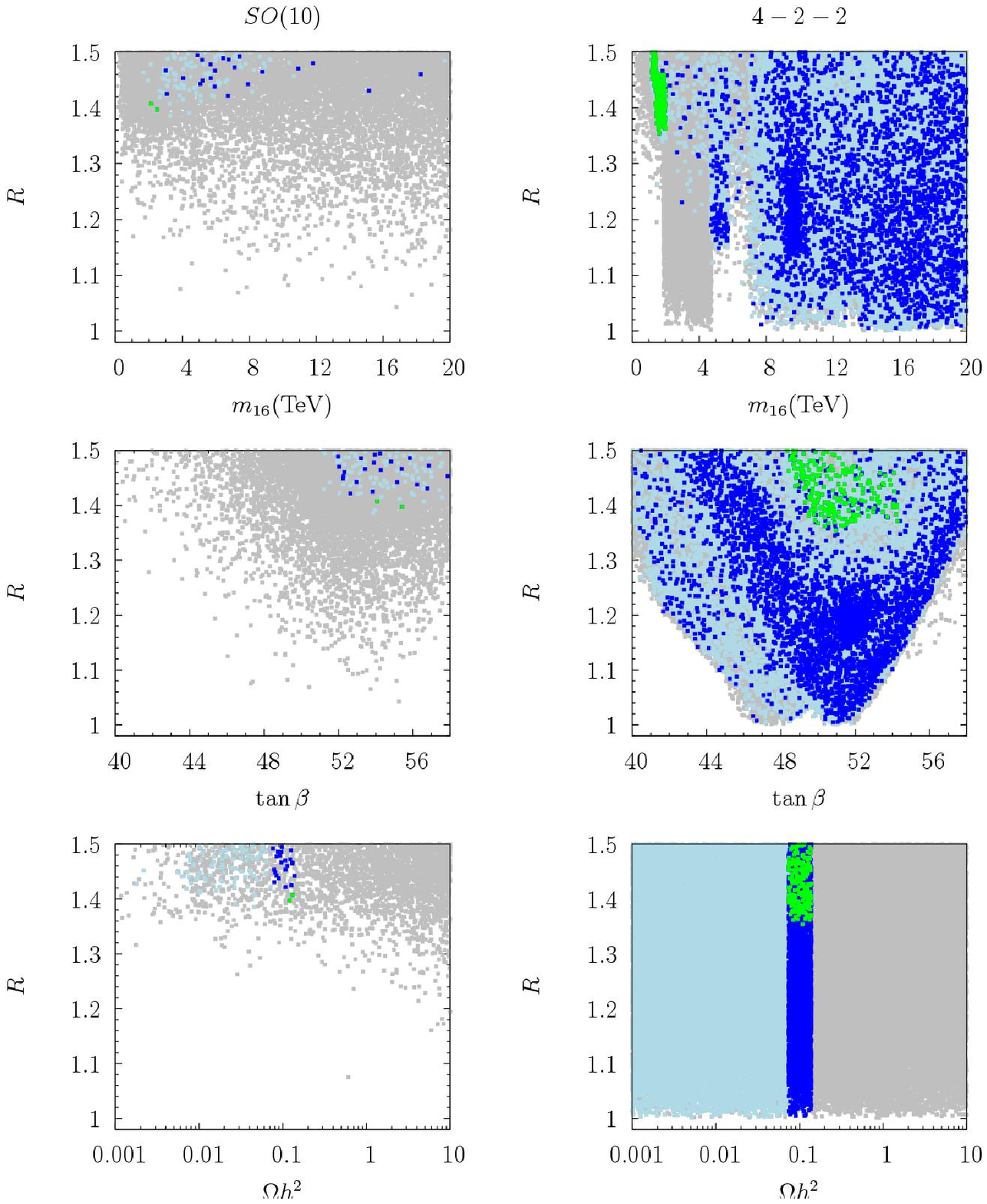}
\caption{Plots in the $(m_{16},R)$, $(\tan\beta,R)$ and $(\Omega
h^2,R)$ planes for SO(10) (left panels) and 4-2-2 (right panels).
Gray points are consistent with REWSB and $\tilde{\chi}^0_{1}$ LSP.
Light blue points satisfy the WMAP upper bound on $\tilde{\chi}^0_1$
abundance and various constraints from colliders ($BR(B_s\rightarrow
\mu^+ \mu^-)$, $BR(b\rightarrow s \gamma)$, and (s)particle mass
bounds). Dark blue points also satisfy the lower bound on
$\tilde{\chi}^0_1$ density. Green points, additionally, satisfy the
constraint from $(g-2)_{\mu}$. \label{fund1}}
\end{figure}

\begin{figure}
\centering
\includegraphics{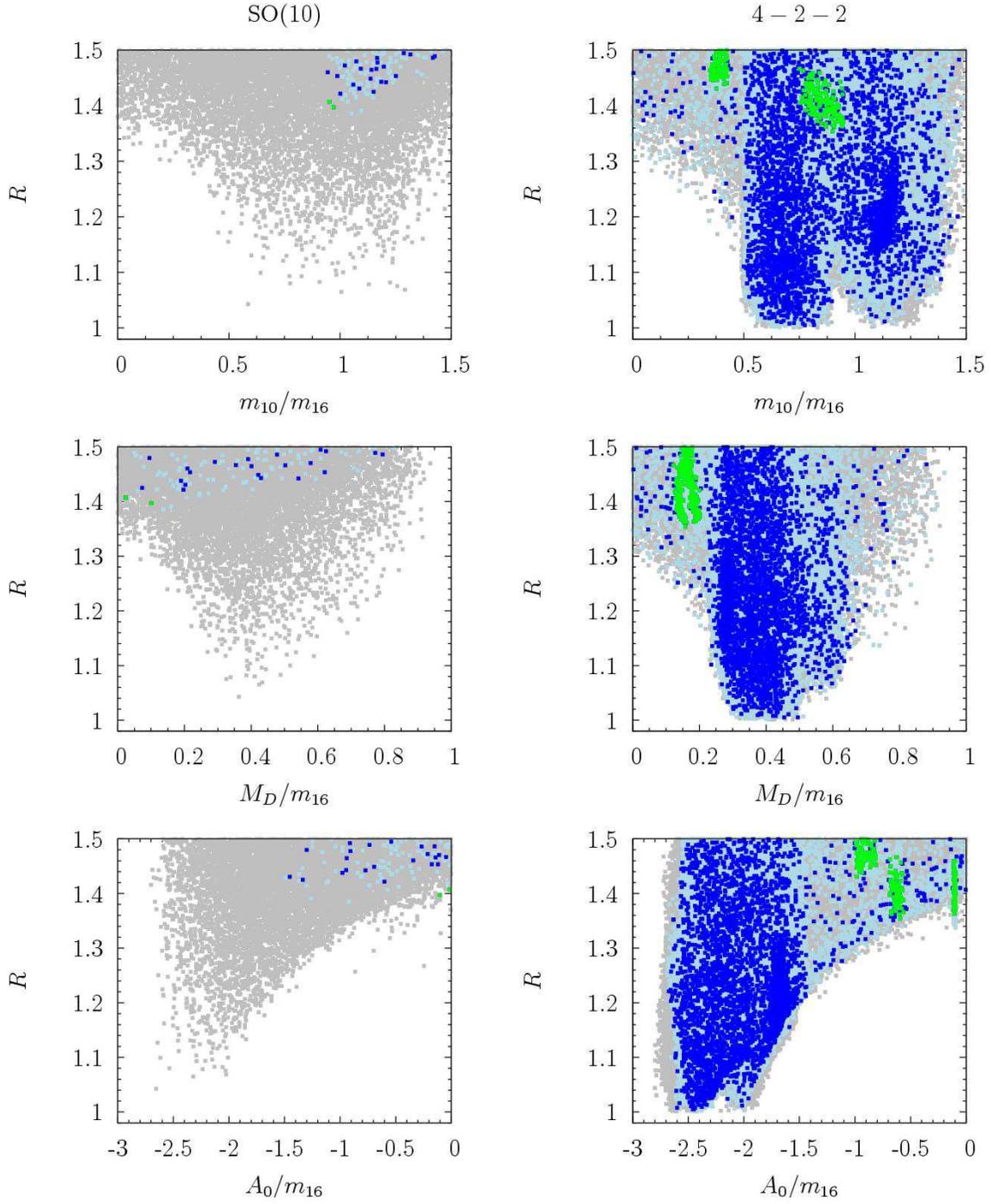}
\caption{Plots in the $(m_{10}/m_{16},R)$, $(M_{D}/m_{16},R)$ and $(A_{0}/m_{16},R)$
planes for SO(10) (left) and 4-2-2 (right). Color coding is the same as
in Figure \ref{fund1}.
\label{fund2}}
\end{figure}

We next present the results of the random scan. We first compare the
$SO(10)$ model with the 4-2-2 model in Figures~\ref{fund1} and
\ref{fund2} following the treatment in \cite{Baer:2008jn}. In
Figure~\ref{fund1} we plot the results in the $(R, m_{16})$,
$(R,\tan\beta)$ and $(\Omega h^2,R)$ planes for $SO(10)$ (left
panel) and 4-2-2 (right panel). All of these points satisfy the
theoretical requirement of REWSB and correspond to a neutralino LSP.
In addition, these points satisfy the various experimental
constraints listed earlier. The light blue points satisfy the
constraints from $BR(B_s\rightarrow \mu^+ \mu^-)$, $BR(b\rightarrow
s \gamma)$, the Higgs, chargino, gluino and stau mass bounds, and
the upper bound on the relic density of dark matter from WMAP. Shown
in dark blue are points that also satisfy the lower bound on
$\tilde{\chi}^0_1$ dark matter abundance. In Figure~\ref{fund2} we
similarly present results in the $(m_{10}/m_{16},R)$,
$(M_{D}/m_{16},R)$ and $(A_{0}/m_{16},R)$ planes for $SO(10)$ (left)
and 4-2-2 (right). It is quite obvious from the results that, as
expected, using just a random scan it is quite difficult to realize
acceptable Yukawa unification in $SO(10)$ consistent with the
experimental constraints. Ref. \cite{Baer:2008jn} employs a modified
scanning algorithm based on Markov Chain Monte Carlo (MCMC) to
search the parameter space more efficiently. It is shown there  that
they show that only the $h$-resonance (light Higgs) channel is
available to bring the neutralino dark matter density in the right
(WMAP) ball park. While this channel does yield acceptable Yukawa
unification consistent with WMAP, it is more or less ruled out by
the lower bound of 114.4 GeV on the the SM Higgs mass.

 In the initial sweep of the $SO(10)$ model around 150,000 points were identified,
consistent with REWSB and the requirement that is a LSP neutralino.
Yukawa unification consistent with the experimental data was found
to be no better than 40\%, even if we ignore the constraint from
$\Delta a_{\mu}$. The 4-2-2 model yields `good' solutions  with
Yukawa unification to better than 10\%. More concentrated searches
around such `good' points have yielded `near perfect' unification.
Such concentrated searches were not performed for the $SO(10)$ model
as they have already been reported in \cite{Baer:2008jn} with the
conclusion that a narrow, almost excluded, light Higgs funnel region
is the only one that is viable from the point of view of Yukawa
unification and dark matter relic density.

We now focus on the 4-2-2 model, which  does much better than the
$SO(10)$ model in terms of Yukawa unification and most of the
experimental constraints, including the WMAP bounds on dark matter
abundance. The constraint from $(g-2)_{\mu}$ is  found to be largely
incompatible with Yukawa unification (Yukawa unification is worse
than 35\% if one insists on $(g-2)_{\mu}$).
 From Figures~\ref{fund1} and
\ref{fund2} we find that  the following parameter values are
preferred:
\begin{eqnarray}
m_{16}\gtrsim 7~{\rm TeV},          \nonumber \\
46\lesssim \tan\beta \lesssim 48~{\rm and}~50 \lesssim \tan\beta \lesssim 52,        \nonumber \\
0.6\lesssim m_{10}/m_{16} \lesssim 0.8~{\rm and}~m_{10}/m_{16}\approx1.1,    \nonumber \\
0.3\lesssim M_{D}/m_{16}\lesssim0.5,     \nonumber \\
A_0\approx -2 m_{16}~{\rm and}~A_0\approx -2.5 m_{16}.
\end{eqnarray}

In Table~\ref{table1} we show a few benchmark points that are
consistent with Yukawa unification. Point 1 displays the spectrum
corresponding to essentially perfect unification ($R=1.00$). Point 2
gives a `light' gluino ($\sim 265$ GeV)  consistent with `good'
unification ($\sim$9\%). Point 3 has the lightest stop (1911 GeV),
again consistent with respectable Yukawa unification ($\sim$7\%).
Note that most of the sparticles are rather heavy as a consequence
of requiring Yukawa unification. Note that for all three benchmark
points the lightest neutralino (LSP) relic abundance is compatible
with the WMAP dark matter bounds. This comes about because of the
relatively small mass splitting between the neutralino (essentially
bino-like) and gluino which leads to efficient co-annihilation
\cite{Profumo:2004wk}.

\begin{figure}
\centering
\includegraphics{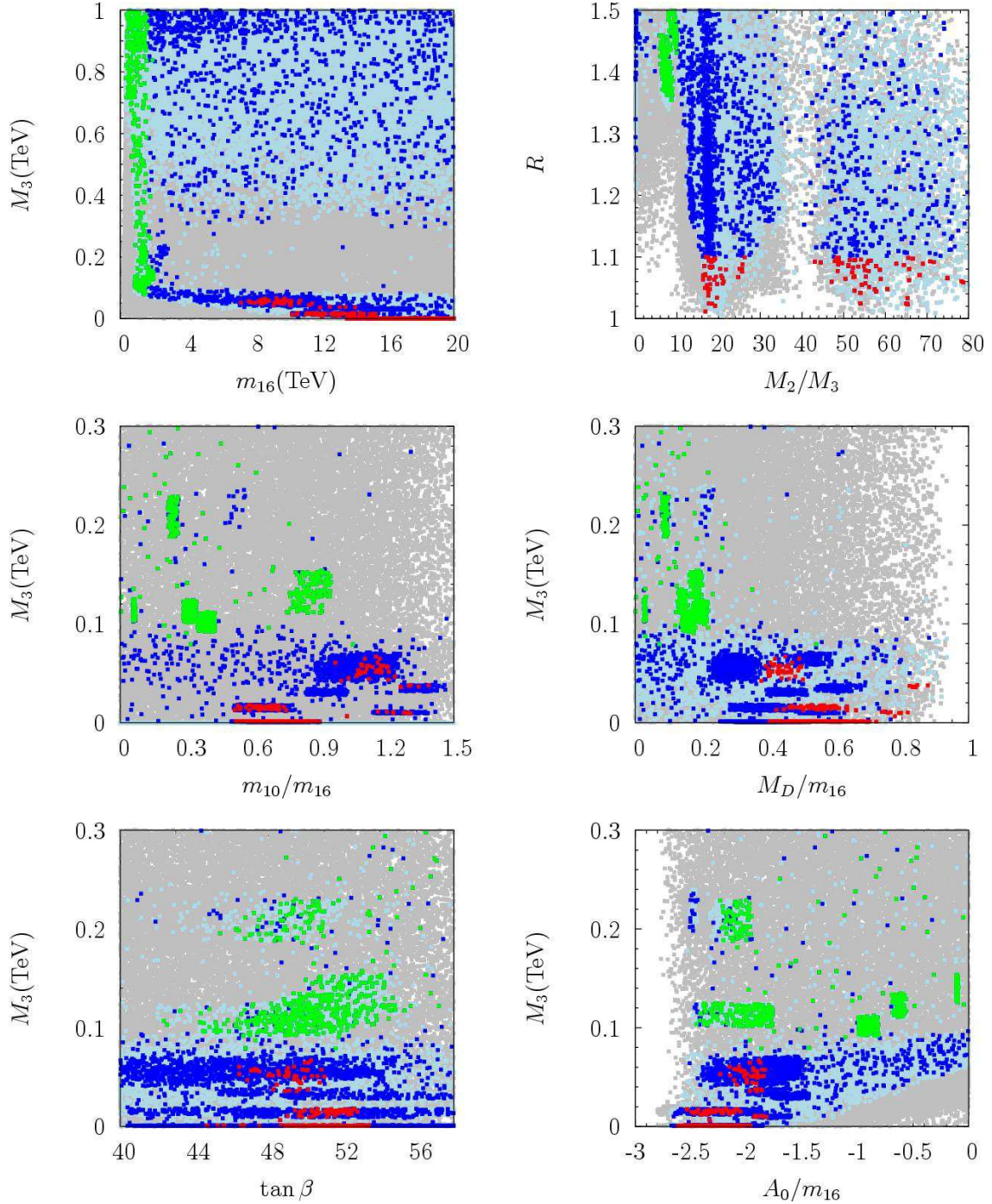}
\caption{Plots in the $(M_3,m_{16})$, $(R,M_2/M_3)$,
$(M_3,m_{10}/m_{16})$, $(M_3,M_D/m_{16})$, $(M_3,\tan\beta)$ and
$(M_3,A_0/m_{16})$ planes for 4-2-2. Gray points are consistent with
REWSB and $\tilde{\chi}^0_{1}$ LSP. Light blue points satisfy the
WMAP upper bound on $\tilde{\chi}^0_1$ abundance and various
constraints from colliders ($BR(B_s\rightarrow \mu^+ \mu^-)$,
$BR(b\rightarrow s \gamma)$, and (s)particle mass bounds). Dark blue
points also satisfy the lower bound on $\tilde{\chi}^0_1$ primordial
abundance. Green points, additionally, satisfy the constraint from
from $(g-2)_{\mu}$. Points in red represent a subset of dark blue
ones that is consistent with 10\% or better Yukawa unification.
\label{fund3}}
\end{figure}

Figure~\ref{fund3} shows plots in the $(M_3,m_{16})$, $(R,M_2/M_3)$,
$(M_3,m_{10}/m_{16})$, $(M_3,M_D/m_{16})$, $(M_3,\tan\beta)$ and
$(M_3,A_0/m_{16})$ planes for the 4-2-2 model. Color coding is
essentially the same as in Figure~\ref{fund1}, except that we now
also show in red points that are consistent with all experimental
constraints (except $(g-2)_{\mu}$) and have Yukawa unification
better than 10\%. It appears that the points with Yukawa unification
seem to favor a non-universal gaugino sector, with $M_2\gtrsim 10
M_3$. This ratio is higher still if we also require these solutions
to satisfy constraints from experiments. This, of course, does not
mean that solutions with $M_2\approx M_3$ do not exist, as the
latter have been reported in \cite{Baer:2008jn}. However, this does
suggest a statistical preference for solutions with a significant
splitting in the gaugino sector.

\begin{figure}
\centering
\includegraphics{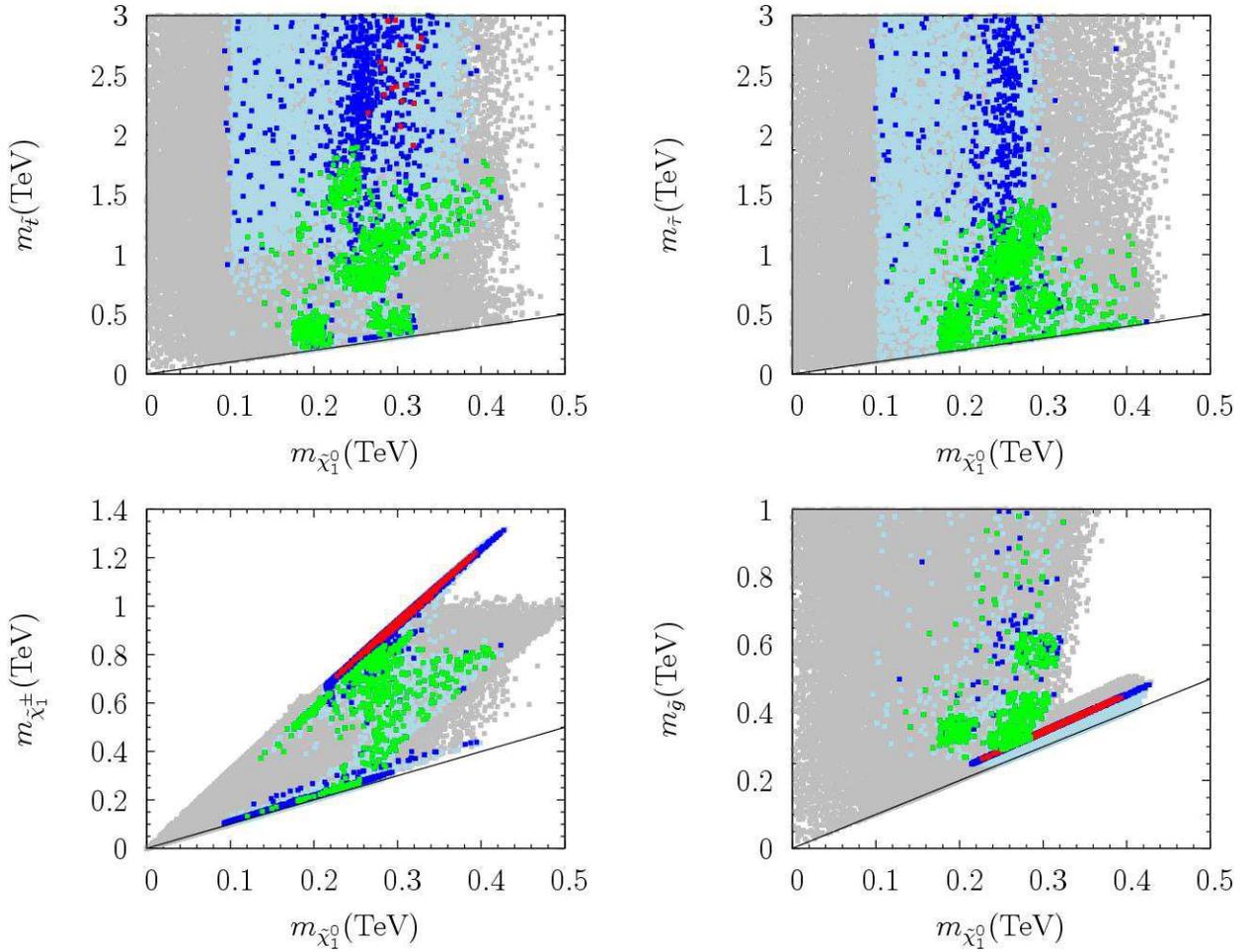}
\caption{Plots in the $(m_{\tilde t},m_{\tilde{\chi}^0_1})$,
$(m_{\tilde\tau},m_{\tilde{\chi}^0_1})$,
$(m_{\tilde{\chi}^{\pm}_1},m_{\tilde{\chi}^0_1})$ and
$(m_{\tilde g},m_{\tilde{\chi}^0_1})$ planes for 4-2-2. Color coding same
as in Figure \ref{fund3}. Also shown is the unit slope line in each plane.
\label{spar1}
}
\end{figure}

The PAMELA experiment has reported an excess in the observed
positron flux with no corresponding anti-proton excess
\cite{Adriani:2008zr}. It may be possible to explain this `excess'
in the context of SUSY with the lightest neutralino as the dark
matter candidate. One explanation invokes a neutralino of mass
around 300 GeV decaying into positrons via `tiny' ($\sim 10^{-13}$)
R-parity violating couplings \cite{Gogoladze:2009kv}. This scenario
is consistent with Yukawa unification as we can see in
Figure~\ref{spar1}.

\begin{figure}
\centering
\includegraphics{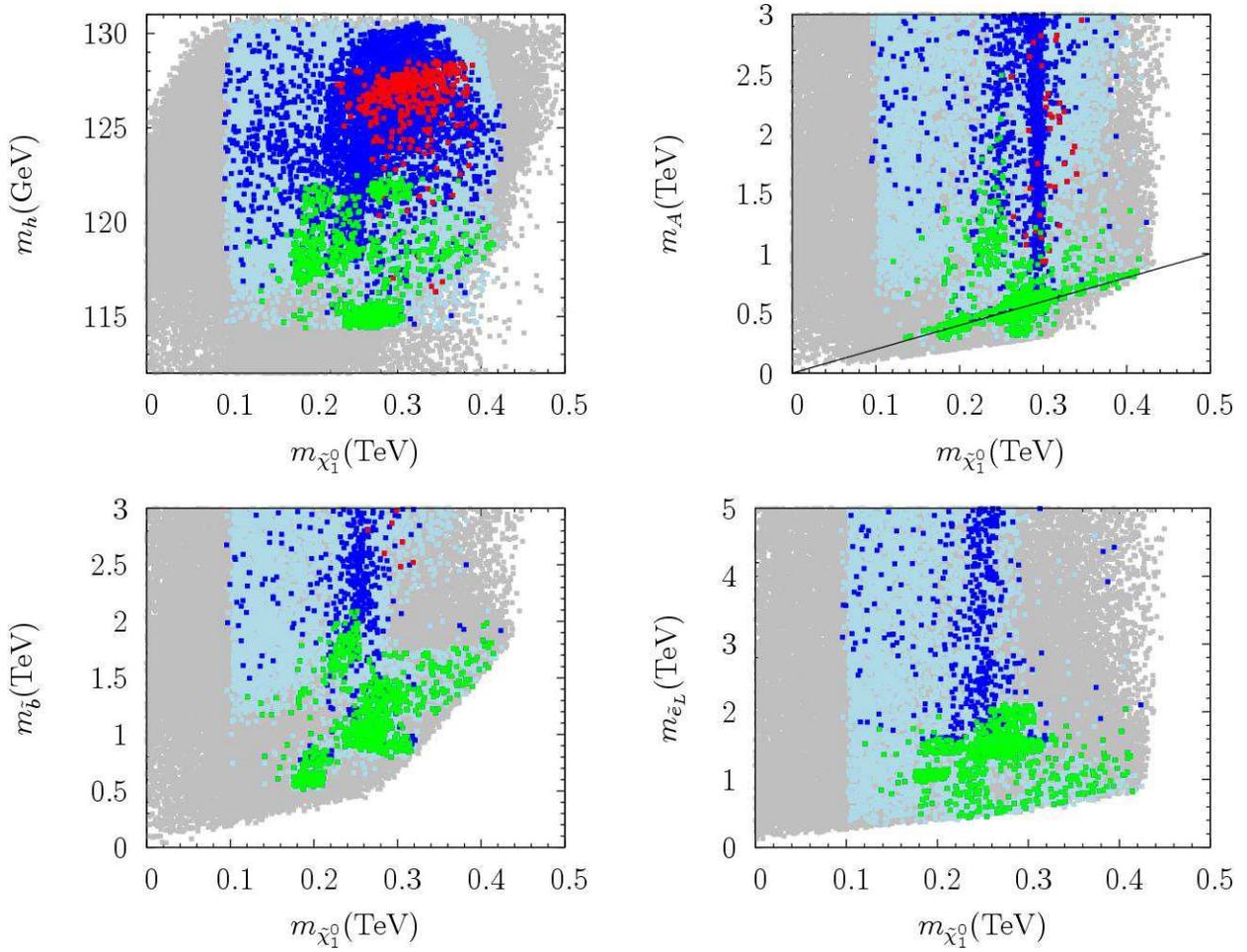}
\caption{Plots in the $(m_{h},m_{\tilde{\chi}^0_1})$,
$(m_{A},m_{\tilde{\chi}^0_1})$,
$(m_{\tilde b},m_{\tilde{\chi}^0_1})$ and
$(m_{\tilde{e}_L},m_{\tilde{\chi}^0_1})$ planes for 4-2-2. Color coding same
as in Figure \ref{fund3}. In the $(m_{A},m_{\tilde{\chi}^0_1})$ case we also show
the line $m_{A}=2 m_{\tilde{\chi}^0_1}$.
\label{spar2}}
\end{figure}

We have stressed that Yukawa unification seems incompatible with the
current experimental bound on $(g-2)_{\mu}$. If we do not insist on
Yukawa unification in 4-2-2, we can find a much lighter MSSM
spectrum, which is consistent with all experimental constraints
(including $(g-2)_{\mu}$). This can be seen from Figures~\ref{spar1}
and \ref{spar2}. In Figure~\ref{spar1} we show plots in the
$(m_{\tilde t},m_{\tilde{\chi}^0_1})$,
$(m_{\tilde\tau},m_{\tilde{\chi}^0_1})$,
$(m_{\tilde{\chi}^{\pm}_1},m_{\tilde{\chi}^0_1})$ and $(m_{\tilde
g},m_{\tilde{\chi}^0_1})$ planes, with the same color coding as in
Figure \ref{fund3}. We also show the unit slope line in each plot,
thus highlighting the stop co-annihilation region, the stau
co-annihilation region, and the mixed bino-wino dark matter region.
In Figure~\ref{spar2} we show similar plots in the
$(m_{h},m_{\tilde{\chi}^0_1})$, $(m_{A},m_{\tilde{\chi}^0_1})$,
$(m_{\tilde b},m_{\tilde{\chi}^0_1})$ and
$(m_{\tilde{e}_L},m_{\tilde{\chi}^0_1})$ planes. We indicate the
A-funnel region with the line $m_A=2 m_{\tilde{\chi}^0_1}$. In
Table~\ref{table2} we present points corresponding to the lightest
spectrum found in our investigation (disregarding Yukawa
unification, but consistent with all experimental constraints).
Points 1 through 5 respectively display the spectrum corresponding
to the lightest chargino (133 GeV), CP-odd Higgs (284 GeV), gluino
(268 GeV), stau (198 GeV) and stop (226 GeV).

\section{Conclusion \label{conclusions}}
The 4-2-2 gauge symmetry, supplemented by left-right symmetry
($C$-parity) captures many attractive features exhibited by the
simplest $SO(10)$ models. One of these features happens, in some
models, to be Yukawa unification. We have shown that by relaxing in
4-2-2 the assumption of universal gaugino masses, the resulting MSSM
models have rather distinctive mass spectra which can be tested at
the LHC. Moreover, the primordial abundance of the lightest
neutralino in this case is consistent with the WMAP dark matter
limits, something which is difficult to achieve in $SO(10)$ with
$t-b-\tau$ Yukawa unification. We have also studied the implications
of relaxing the Yukawa unification condition and identified several
additional benchmark points which also can be explored at the LHC.
Finally, we wish to note the recent observation that the little
hierarchy problem can be largely resolved in the 4-2-2 framework
\cite{lit-h}. The implication of this for sparticle spectroscopy
will be discussed elsewhere.

\section*{Acknowledgments}
We thank Azar Mustafayev and Masnoor Rehman for helpful discussions. This work
is supported in part by the DOE Grant \# DE-FG02-91ER40626 (I.G., R.K. and Q.S.),
GNSF grant 07\_462\_4-270 (I.G.), and by Bartol Research Institute (R.K.).

\newpage

\begin{table}[t]
\centering
\begin{tabular}{lcccc}
\hline
\hline
& Point 1 & Point 2 & Point 3      \\
\hline
$m_{16}$ & 14110  & 8429   & 13124        \\
$M_{2} $ &  832.03  & 1020.2  & 689.4       \\
$M_{3} $ &  0.7945  & 60.542  & 9.6261        \\
$\tan\beta$ & 50.82   & 46.41 & 51.17        \\
$M_{D}/m_{16}$ &  0.4543  & 0.5595   & 0.3323    \\
$m_{10}/m_{16}$ &  0.7741   & 1.1584  & 1.3048   \\
$A_{0}/m_{16} $ &  -2.4487  & -2.1527 & -1.8226      \\
\hline
$m_h$          & 123    & 126    & 127           \\
$m_H$          & 7569   & 2163   & 9882            \\
$m_A$          & 7520   & 2150   & 9818          \\
$m_{H^{\pm}}$  & 7571   & 2175   & 9883          \\
\hline
$m_{\tilde{\chi}^{\pm}_{1,2}}$
& 887,13869 & 975,4047  & 712,3750     \\
$m_{\tilde{\chi}^0_{1,2}}$ &  283,885 & 319,974  & 228,712      \\
$m_{\tilde{\chi}^0_{3,4}}$ &  13879,13879 & 4049,4049  & 3784,3785       \\
$m_{\tilde{g}}$ & 325 & 365 & 265    \\
\hline $m_{ \tilde{u}_{L,R}}$
& 14126,13916 & 8435,8361  & 13140,12841      \\
$m_{\tilde{t}_{1,2}}$
& 5337,5726 & 1911,2640  & 4931,5310    \\
\hline $m_{ \tilde{d}_{L,R}}$
& 14126,14203 & 8435,8455   & 13141,13249     \\
$m_{\tilde{b}_{1,2}}$
& 5237,5653 & 2521,2767  & 4115,5146     \\
\hline
$m_{\tilde{\nu}_{1}}$
& 13988 & 8409   & 12926       \\
$m_{\tilde{\nu}_{3}}$
& 10598 & 6577   & 9535        \\
\hline
$m_{ \tilde{e}_{L,R}}$
& 13988,14376  & 8408,8514  & 12926,13500     \\
$m_{\tilde{\tau}_{1,2}}$
& 6412,10581  & 4270,6573   & 5580,9559      \\
\hline
$\mu$   & 14100  &  4110  & 3840     \\
\hline
$\Omega_{LSP}h^2$ &  0.09 & 0.112   & 0.116    \\
\hline
$R$ & 1.00 & 1.07  & 1.09   \\
\hline
\hline
\end{tabular}
\caption{ Sparticle and Higgs masses in 4-2-2 model (in units of
GeV), with $m_t=172.6$ GeV and $\mu>0$. Point 1 corresponds to exact
Yukawa unification ($R=1.00$) while  points 2 (3) shows the spectrum
corresponding to the lightest stop (gluino) with Yukawa unification
of 10\% or better.Note that in each case gluino co-annihilation
plays as essential role.} \label{table1}
\end{table}

\begin{table}[t]
\centering
\begin{tabular}{lcccccc}
\hline
\hline
& Point 1 & Point 2 & Point 3 & Point 4 & Point 5      \\
\hline
$m_{16}$ & 1529.4   & 1038.5   & 1402.5 & 958.4 & 1469.8   \\
$M_{2} $ &  158.0  & 630.1   & 736.5 & 607.6 & 630.1  \\
$M_{3} $ &  467.9  & 122.0   & 79.43 & 117.8 & 103.6  \\
$\tan\beta$ & 56.4   & 57.2 & 46.8 & 54.5 & 46.2  \\
$M_{D}/m_{16}$ &  0.2185  & 0.2085   & 0.0721 & 0.1732 & 0.0276  \\
$m_{10}/m_{16}$ &  0.459  & 0.339   & 0.317 & 0.291 & 0.059  \\
$A_{10}/m_{16} $ &  -1.485  & -1.976   & -1.434 & -2.063 & -2.45   \\
\hline
$m_h$          & 119   & 118   & 117   & 119  &  120     \\
$m_H$          & 940   & 284   & 448   & 297  &  468      \\
$m_A$          & 934   & 284   & 445   & 299  &  472     \\
$m_{H^{\pm}}$  & 946   & 302   & 458   & 315  &  491     \\
\hline
$m_{\tilde{\chi}^{\pm}_{1,2}}$
& 133,1545 & 526,1070 & 620,1238 & 505,1015 & 541,1676  \\
$m_{\tilde{\chi}^0_{1,2}}$ &  121,132 & 186,526 & 208,620 & 178,504 & 186,539   \\
$m_{\tilde{\chi}^0_{3,4}}$ &  1543,1543 & 1065,1068 & 1232,1236 & 1010,1014 & 1675,1676   \\
$m_{\tilde{g}}$ & 1176 & 368 & 268 & 354 & 335  \\
\hline $m_{ \tilde{u}_{L,R}}$
& 1784,1784 & 1135,1064 & 1475,1406 & 1056,986 & 1527,1481   \\
$m_{\tilde{t}_{1,2}}$
& 1148,1392 & 409,764 & 777,1080 & 319,709 & 226,857  \\
\hline $m_{ \tilde{d}_{L,R}}$
& 1785,1790 & 1138,1066 & 1477,1404 & 1059,987 & 1530,1479  \\
$m_{\tilde{b}_{1,2}}$
& 1331,1497 & 613,799 & 972,1115 & 560,743 & 739,987  \\
\hline
$m_{\tilde{\nu}_{1}}$
& 1526 & 1115  & 1482 & 1036 & 1526  \\
$m_{\tilde{\nu}_{3}}$
& 1249 & 852  & 1293 & 797 & 1191  \\
\hline
$m_{ \tilde{e}_{L,R}}$
& 1528,1540  & 1118,1055 & 1483,1413 & 1039,974 & 1527,1478  \\
$m_{\tilde{\tau}_{1,2}}$
& 878,1261  & 200,864 & 960,1299 & 198,808 & 586,1192   \\
\hline
$\mu$ & 1555 & 1077 & 1247 & 1020 & 1685  \\
\hline
$\Omega_{LSP}h^2$ &  0.079 & 0.076  & 0.074 & 0.114 & 0.127  \\
\hline
\hline
\end{tabular}
\caption{ Sparticle and Higgs masses (in units of GeV), with
$m_t=172.6$ GeV and $\mu>0$. Points 1 through 5 correspond to the
lightest chargino, CP-odd Higgs, gluino, stau and stop for the 4-2-2
parameter space given in Eq. (\ref{ppp1}). Note that these points
are not consistent with Yukawa unification ($R>2.3$), but they
satisfy all experimental constraints including the one from
$(g-2)_{\mu}$.} \label{table2}
\end{table}

\end{document}